\documentclass[conference]{IEEEtran}
\IEEEoverridecommandlockouts
\usepackage{cite}
\usepackage{standalone}
\usepackage{amsmath,amssymb,amsfonts}
\usepackage{systeme,mathtools}
\newcommand\myeq{\stackrel{\mathclap{\normalfont\mbox{(a)}}}{=}}
\newcommand\myeqb{\stackrel{\mathclap{\normalfont\mbox{(b)}}}{=}}

\usepackage{accents}
\newlength{\dhatheight}

\usepackage[dvipsnames]{xcolor}
\usepackage{algorithmic}
\usepackage{graphicx}
\usepackage{textcomp}
\usepackage{xcolor}
\def\BibTeX{{\rm B\kern-.05em{\sc i\kern-.025em b}\kern-.08em
    T\kern-.1667em\lower.7ex\hbox{E}\kern-.125emX}}
\usepackage{tikz}
\usepackage{graphicx}
    \usepackage{pgfplots}
\pgfplotsset{compat=newest}

\begin{document}

\title{Joint Sequential Fronthaul Quantization and Hardware Complexity Reduction in Uplink Cell-Free Massive MIMO Networks \\
 \thanks{This work is supported by the European Union's
Horizon 2020 research and innovation program under grant
agreements: 101013425 (REINDEER) and 101017171 (MARSAL) and by Research Foundation – Flanders (FWO)
project number G0C0623N.

The resources and services used in this work were provided by the VSC (Flemish Supercomputer Center), funded by the Research Foundation - Flanders (FWO) and the Flemish Government.

© 2024 IEEE.  Personal use of this material is permitted.  Permission from IEEE must be obtained for all other uses, in any current or future media, including reprinting/republishing this material for advertising or promotional purposes, creating new collective works, for resale or redistribution to servers or lists, or reuse of any copyrighted component of this work in other works.
}
}
\author{\IEEEauthorblockN{Vida Ranjbar\IEEEauthorrefmark{1}, Robbert Beerten\IEEEauthorrefmark{1}, Marc Moonen\IEEEauthorrefmark{1}, Sofie Pollin\IEEEauthorrefmark{1}\IEEEauthorrefmark{2}}
\IEEEauthorblockA{\IEEEauthorrefmark{1}\textit{Department of Electrical Engineering, KU Leuven, Belgium} \\
\IEEEauthorrefmark{2}\textit{IMEC, Kapeldreef 75, 3001 Leuven, Belgium} \\
Corresponding Author: \{vida.ranjbar\}@kuleuven.be}}

\maketitle

\begin{abstract}
Fronthaul quantization causes a significant
distortion in cell-free massive MIMO networks. Due to the limited capacity of fronthaul links, information exchange among access points (APs) must be quantized significantly. Furthermore, the complexity of the multiplication operation in the base-band processing unit increases with the number of bits of the operands. Thus, quantizing the APs' signal vector reduces the complexity of signal estimation in the base-band processing unit. Most recent works consider the direct quantization of the received signal vectors at each AP without any pre-processing. However, the signal vectors received at different APs are correlated mutually (inter-AP correlation) and also have correlated dimensions (intra-AP correlation). Hence, cooperative quantization of APs fronthaul can help to efficiently use the quantization bits at each AP and further reduce the distortion imposed on the quantized vector at the APs. This paper considers a daisy chain fronthaul and three different processing sequences at each AP. We show that 1) de-correlating the received signal vector at each AP from the corresponding vectors of the previous APs (inter-AP de-correlation) and 2) de-correlating the dimensions of the received signal vector at each AP (intra-AP de-correlation) before quantization helps to use the quantization bits at each AP more efficiently than directly quantizing the received signal vector without any pre-processing and consequently, improves the bit error rate (BER) and normalized mean square error (NMSE) of users signal estimation.
\end{abstract}


\section{Introduction}
Massive multiple-input-multiple-output (MIMO) networks are very well known for their ability to spatially multiplex users using a large number of antennas. Spatial multiplexing enables the users to use the same time and frequency resources and hence improves users' spectral efficiency. In Cell-free massive MIMO (CFmMIMO), the antennas are distributed among multiple distributed access points (APs), which are coordinated by a central processing unit (CPU) \cite{cellfreevssmallcell, bjornsonmaking, bjornsonscalable}. 
The APs cooperate to serve the users effectively. 
Such a paradigm can further improve the spectral efficiency of the users by exploiting the spatial diversity of the APs. Distributing antennas in multiple small APs helps to alleviate the adverse effect of large-scale fading, such as path-loss and shadowing, on users' channel gain \cite{cellfreevssmallcell} compared to the collocated massive MIMO network. However, this advantage comes with a lot of challenges. One of the challenges is the efficient usage of the limited capacity of the fronthaul links connecting APs to each other or to the CPU \cite{bashar_EE_uq_2019, Bashar_2021_uniform_q}. The authors of \cite{bashar_EE_uq_2019} considered the quantization of the pilot and data vector in the uplink to be sent through the fronthaul link to the CPU. They considered three cases based on where the channel is estimated and used for users' signal estimation. In addition, in \cite{masoumiperformance}, authors considered fronthaul rate allocation to corresponding signals of different users. 
In \cite{adap_bit_svd}, the authors considered a CFmMIMO network in which the APs are connected to the CPU in a star topology, and each AP first reduces the number of streams that it sends to the CPU using singular value decomposition (SVD) of the received signal vector and then allocate bits to the streams to maximize sum signal to noise ratio (SNR) of the streams.
However, in all the works mentioned above, it is assumed that the APs are connected to the CPU in a star topology, and each AP quantizes its received vector individually and in isolation from other APs.

Besides the limited capacity fronthaul links, the hardware size and complexity of the base-band processing unit in each AP is of great importance as the APs in a CFmMIMO are supposed to be cheap entities with low hardware complexity. Thus, a significant effort should be invested regarding efficient low-bit quantization of the received signal vectors to meet the capacity constraint of the fronthaul link and hardware complexity constraint of the base-band processing units.
\subsection{Motivation}
Besides star topology, daisy chain or sequential topology has also been considered recently \cite{shaik2020, Shaik2021, ke_Dmimo_kalman, ranjbar2022} in the context of cell-free massive MIMO networks. In such a topology, each AP refines user signal estimates in the uplink based on the information received from the previous AP over a capacity-limited fronthaul link in the chain. Hence, the efficient usage of fronthaul capacity is important. However, the authors in \cite{shaik2020, Shaik2021} don't consider the limited capacity fronthaul constraint impact on the performance. The authors in \cite{ranjbar2022} investigate the convergence behavior of recursive least squares algorithms under limited capacity fronthaul links with quantizers operating individually. However, in this work, we would like to consider the impact of \textbf{\textit{inter-AP}} information on quantization.

To meet the fronthaul and hardware requirement, it is possible to quantize the raw received signal vector element-wise, which is not recommended as the elements of the raw received signal vector at each AP are correlated. A better approach is to first de-correlate the dimensions of the raw received signal vector at each AP and then quantize it element-wise. However, the received signal vectors among APs are also usually correlated, conditioned on the local channel state information (CSI). Hence, in a third approach, APs can use the information received from the previous AP in the chain to de-correlate their received signal vector from the signal vector of the previous APs in the chain before quantization. This will allow the APs to use the quantization bits efficiently for quantizing the received signal vector. 
The APs then use the quantized vector to refine user signal estimates. 

To clarify the impact of de-correlation on the efficient usage of the quantization bits, 
consider the following toy example.
Suppose we want to estimate a random variable $x$ based on the realization of two other random variables, $y_1$ and $y_2$. Assume that $x$, $y_1$ and $y_2$ have zero mean. We consider the linear minimum mean square error (LMMSE) estimate of one of them with respect to others. Hence, if $\mathbb{E}\{x\begin{bmatrix}
    y_1\\y_2
\end{bmatrix}^{\text{H}}\}\neq\mathbf{0}$, LMMSE estimate of $x$ with respect to $y_1$ and $y_2$ is as follows:
\begin{equation}
    \hat{x}=v_1y_1+v_2y_2.
    \label{mmse_toy}
\end{equation}
where $v_i=\mathbb{E}\{xy_i^{\text{H}}\}\mathbb{E}\{y_iy_i^{\text{H}}\}^{-1}, i\in\{1,2\}$ \cite{dougherty1999}.
Similarly, if $y_1$ and $y_2$ are correlated, i.e. $\mathbb{E}\{y_1y_2^{\text{H}}\}\neq0$, we calculate LMMSE estimate of $y_2$ with respect to $y_1$ as follows:
\begin{equation}
    \hat{y}_2=by_1,
    \label{mmse_obs}
\end{equation} 
then, it follows that:
\begin{equation}
    y_2=\hat{y}_2+\tilde{y}_2,
    \label{est_err}   
\end{equation}
where $\tilde{y}_2$ is the estimation error.
Based on the orthogonality principle of the LMMSE estimate:
\begin{equation}
    \mathbb{E}\{y_1\tilde{y}^{\text{H}}_2\}=0,
\end{equation}
or in general, the estimation error $\tilde{y}_2$ is uncorrelated to any linear function of $y_1$.
Based on (\ref{mmse_toy}), (\ref{mmse_obs}) and (\ref{est_err}), we have:
\begin{equation}
    \hat{x}=v_1y_1+v_2(\hat{y}_2+\tilde{y}_2)=(v_1+v_2b)y_1+v_2\tilde{y}_2.
\end{equation}
Therefore, the knowledge of $\tilde{y}_2$ (instead of $y_2$) is enough to estimate $x$. Now, consider that there is a quantization step for both observations before the estimation of $x$. As 1) knowledge of $\tilde{y}_2$ is enough for estimating $x$, and 2) based on the orthogonality principle of LMMSE estimator, the variance of the $\tilde{y}_2$ is smaller than ${y}_2$, i.e., $\mathbb{E}\{\tilde{y}_2\tilde{y}_2^{\text{H}}\}<\mathbb{E}\{{y}_2{y}_2^{\text{H}}\}$, the error of quantizing $\tilde{y}_2$ is smaller than ${y}_2$ while using uniform quantization with a certain number of bits. Therefore, it is recommended to de-correlate $y_2$ from $y_1$ and then quantize $y_1$ and $\tilde{y}_2$.
\subsection{Notation}
    We denote vectors and matrices with boldface lower-case and upper-case letters, respectively. Transpose and conjugate transpose operations are denoted by superscripts $^{\text{T}}$ and $^{\text{H}}$, respectively. A circularly symmetric complex Gaussian distribution with covariance matrix $\mathbf{X}$ is represented as $\mathcal{C}\mathcal{N}(0, \mathbf{X})$. Symbol $\mathbb{E}\{\mathbf{x}\}$ denotes the mean of $\mathbf{x}$. $\operatorname{Re}(\mathbf{x})$ and $\operatorname{Im}(\mathbf{x})$ denote the elemen-wise real and imaginary part of $\mathbf{x}$, respectively. $diag(\mathbf{x})$ is a diagonal matrix with the same diagonal elements as the elements of vector $\mathbf{x}$.
\subsection{Contribution}
This paper uses joint fronthaul quantization and hardware complexity reduction in a cell-free massive MIMO network with sequential fronthaul. In practical scenarios, the number of bits at each AP to quantize the received signal vector is limited by 1) fronthaul capacity between the APs and 2) hardware complexity in the base-band processing unit in each AP. We consider three Options for processing sequence at each AP, as shown in Fig. \ref{3cases}. We show that AP $l$ can more efficiently use the quantization bits to quantize its received signal vector in {\bf Option $1$} where it takes the information sent from AP $l-1$ into account \textbf{before quantization}. In {\bf Option $1$}, AP $l$ quantizes a vector that is not only de-correlated from the corresponding vectors in previous APs (\textbf{\textit{inter-AP}} de-correlation) but also has de-correlated dimensions (\textbf{\textit{intra-AP}} de-correlation). We compared {\bf Option $1$} with 1) {\bf Option $2$} where AP $l$ only considers the de-correlation between the dimensions of its local received signal vector (\textbf{\textit{intra-AP}} de-correlation only) \textbf{before quantization} and 2) {\bf Option $3$} where AP $l$ directly quantizes the received signal vector without any \textbf{\textit{inter-AP}} or \textbf{\textit{intra-AP}} de-correlation (no de-correlation before quantization). The simulation results show the superiority of {\bf Option $1$} over the other two options.     

 \subsection{Outline}
The rest of the paper is organized as follows. In Section \ref{sec2}, we introduce the system model. In Section \ref{sec3}, we consider the uplink signal estimation using distributed MMSE processing. We introduce the concept of dithering and how it facilitates the analysis of the quantization noise. In section \ref{sec4}, we show the numerical result, and finally, section \ref{sec5} concludes the paper.
\section{System Model}\label{sec2}
We consider a cell-free massive MIMO network with $L$ APs connected in a daisy chain topology, each with $N$ antennas serving $K>N$ users in the uplink.
We assume the channel between AP $l$ and user $k$ is a circularly symmetric complex Gaussian random vector, denoted by  $\mathbf{h}_{kl}\in \mathcal{C}\mathcal{N}(\mathbf{0},\mathbf{R}_{\mathbf{h}_{kl}})$, also called correlated Rayleigh fading channel \cite{massivemimobook}.
The large-scale fading coefficient is $\mathbf{\beta}_{kl}=\text{trace}({\mathbf{R}_{\mathbf{h}_{kl}}})/N$.
The received signal vector at AP $l$ is as follows:
\begin{equation}
    \mathbf{y}_l=\mathbf{H}_l\mathbf{s}+\mathbf{n}_l,
\end{equation}
where $\mathbf{H}_l=\begin{bmatrix}
    \mathbf{h}_{1l}&&\hdots&&\mathbf{h}_{Kl}
\end{bmatrix}$ and $\mathbf{n}_l\sim\mathcal{C}\mathcal{N}(0,\sigma^2\mathbf{I}_N)$ is the noise vector at AP $l$. We assume vector   
$\mathbf{s}$ with $\mathbb{E}\{\mathbf{s}\mathbf{s}^H\}=p\mathbf{I}_K$ as the user transmitted signal vector. 
We assume a block fading model channel with $B_c$ as coherence bandwidth and $T_c$ as coherence time. The transmitted signal bandwidth is $B$. The channel matrix will remain constant for $\tau_c=T_cB_c$ samples. Out of $\tau_c$ samples, $\tau_d$ samples are used for uplink data transmission. We assume perfect CSI at the APs.
\section{Uplink processing and users' signal estimation} \label{sec3}
This section considers the sequential estimation of users' signal in the uplink using distributed MMSE processing \cite{ke_Dmimo_kalman, Shaik2021}. In this section, we elaborate mathematically on {\bf Option $1$} processing sequence as shown in Fig. \ref{3cases}.
 As the signal estimation is recursive, i.e., each AP's estimate depends on the previous AP's estimate in the chain to update user signal estimates, we start from AP $1$.
 AP $1$ estimates users' signal as follows:
 \begin{equation}     
 \hat{\mathbf{s}}_1=\mathbf{V}_1{f}_1({{\mathbf{y}}}_1),
 \end{equation}
 Where $\mathbf{V}_1\in \mathbb{C}^{K\times r}$ contains local combining vector, $f_1: \mathbb{C}^N\rightarrow \mathbb{C}^r$ such that $r=min(N,K)$ is the function that includes the de-correlation pre-processing and quantization.
 It will be shown in Subsection \ref{3B} that $f_1$ is a linear function of ${\mathbf{y}}_1$ described as:
 \begin{equation}
     {f}_1({{\mathbf{y}}}_1)=\mathbf{A}_1{\mathbf{y}}_1+\mathbf{b}_1,
 \end{equation}
 where $\mathbf{A}_1 \in\mathbb{C}^{r\times N}$ is a deterministic matrix
and $\mathbf{b}_1\in\mathbb{C}^{r\times 1}$ is a random vector. Vector $\mathbf{b}_1$ results from the de-correlation, dithering, and quantization, elaborated in Subsection {\ref{3B}}.
\subsection{Received signal vector processing at AP l}\label{3A}
Each AP $l$ receives from AP $l-1$: 
\begin{itemize}
    \item For each sample in a coherence block, $K$ user signal estimates, i.e. vector $\hat{\mathbf{s}}_{l-1}$, 
    \item And once per coherence block, the user signal estimation error covariance matrix ${\mathbf{C}}_{l-1}\in \mathbb{C}^{K\times K}$ defined as,
\begin{equation}
    {\mathbf{C}}_{l-1}=\mathbb{E}\{\tilde{\mathbf{s}}_{l-1}\tilde{\mathbf{s}}_{l-1}^H|\mathbf{H}_l\},
    \label{err_cov_mat}
\end{equation}
\end{itemize} 
where 
\begin{equation}
    \tilde{\mathbf{s}}_{l-1}=\mathbf{s}-\hat{\mathbf{s}}_{l-1},
    \label{eq_signal_est_err}
\end{equation}
According to (\ref{eq_signal_est_err}), AP $l$ must refine the user signal estimates of AP ${l-1}$ by estimating the error $\tilde{\mathbf{s}}_{l-1}$ on that previous estimate $\hat{\mathbf{s}}_{l-1}$. To do so, AP $l$ first subtracts the LMMSE estimate of its received signal vector $\mathbf{y}_l$ based on $f_i({\mathbf{y}}_i), i\in\{1,\hdots,l-1\}$ from $\mathbf{y}_l$. It can be shown that this LMMSE estimate can be calculated using only the local channel matrix at AP $l$ and the user signal estimates: 
\begin{equation}
\begin{aligned}    \mathcal{G}_l({\mathbf{y}}_l)&={\mathbf{y}}_l-\text{LMMSE}\{{\mathbf{y}}_l|f_1({\mathbf{y}}_1),\hdots,f_{l-1}({\mathbf{y}}_{l-1})\}\\&\myeq{\mathbf{y}}_l-\mathbf{H}_l\hat{\mathbf{s}}_{l-1}={\mathbf{H}}_l(\mathbf{s}-\hat{\mathbf{s}}_{l-1})+\mathbf{n}_l\\&\myeqb{\mathbf{H}}_l\tilde{\mathbf{s}}_{l-1}+\mathbf{n}_l,
    \label{eq_mmse}
    \end{aligned}
\end{equation}
where $\myeq$ is due to the following fact:
\begin{equation}
    \text{LMMSE}\{{\mathbf{y}}_l|f_1({\mathbf{y}}_1),\hdots,f_{l-1}({\mathbf{y}}_{l-1})\}=\mathbf{H}_l\hat{\mathbf{s}}_{l-1}.
    \label{mmsey}
\end{equation}
The proof is omitted due to space limitations. The interested reader is referred to \cite{Shaik2021}. This LMMSE estimate is fully based on the user signal previous estimates, so there is no new information in it. Hence, it should be removed so that the quantization bits can be used for the remaining information in $\mathbf{y}_l$.
Furthermore, (\ref{eq_signal_est_err}) proves $\myeqb$ in (\ref{eq_mmse}). 

Using the orthogonality principle of the LMMSE estimator:
\begin{equation}
    \mathbb{E}\{f_{i}({\mathbf{y}}_i)\mathcal{G}_l({\mathbf{y}}_l)^{\text{H}}\}=0, \forall i\in\{1,\hdots,l-1\}.
\end{equation}

The process in (\ref{eq_mmse}) is called \textbf{\textit{inter-AP}} de-correlation. However, the dimensions of  $\mathcal{G}_l({\mathbf{y}}_l)$ may still be correlated. To use the quantization bits even more efficiently, AP $l$ also de-correlates the dimensions of $\mathcal{G}_l({\mathbf{y}}_l)$ via the well-known PCA method. The resulting vector is then quantized element-wise.
To use the PCA method, AP $l$ computes the $r$ eigenvectors with the largest corresponding eigenvalues of the 
covariance matrix of the \textbf{\textit{inter-AP}} de-correlated received signal vector, i.e.,  $\mathbf{R}_{\mathcal{G}_l({\mathbf{y}}_l)}=\mathbb{E}\{\mathcal{G}_l({\mathbf{y}}_l)\mathcal{G}_l({\mathbf{y}}_l)^{\text{H}}|\mathbf{H}_l\}$. According to (\ref{eq_mmse}), the SVD decomposition of $\mathbf{R}_{\mathcal{G}_l({\mathbf{y}}_l)}$ is as follows: 
\begin{equation}
\begin{aligned}
\mathbf{R}_{\mathcal{G}_l({\mathbf{y}}_l)}&=\mathbb{E}\{\mathcal{G}_l({\mathbf{y}}_l)\mathcal{G}_l({\mathbf{y}}_l)^H|\mathbf{H}_l\}\\ 
&=\mathbf{H}_l\mathbb{E}\{\tilde{\mathbf{s}}_{l-1}\tilde{\mathbf{s}}_{l-1}^{\text{H}}|\mathbf{H}_l\}\mathbf{H}_l^{\text{H}}+\sigma^2\mathbf{I}_N\\&\myeq\mathbf{H}_l\mathbf{C}_{l-1}\mathbf{H}_l^{\text{H}}+\sigma^2\mathbf{I}_N\\&=\mathbf{U}_l\mathbf{\Sigma}_l\mathbf{U}_l^{\text{H}},
\label{svd}
    \end{aligned}
\end{equation}
where $\myeq$ is a result of (\ref{err_cov_mat}). Matrix $\mathbf{U}_l$ contains the eigenvectors of $\mathbf{R}_{\mathcal{G}_l({\mathbf{y}}_l)}$. Assuming that eigenvalues are sorted in descending order, we select the $r$ most prominent eigenvectors (corresponding to the largest eigenvalues), as follows:
\begin{equation}
    \mathbf{A}_l=\mathbf{U}_{l[:,[1:r]]}.
\end{equation}
Matrix $\mathbf{A}_l$ then projects  $\mathcal{G}_l({\mathbf{y}}_l)$ onto the subspace spanned by these eigenvectors.
The PCA processed version of $\mathcal{G}_l({{\mathbf{y}}}_l)$ is:
\begin{equation} 
\mathcal{P}_l(\mathcal{G}_l({{\mathbf{y}}}_l))=\mathbf{A}_l^{\text{H}}\mathcal{G}_l({{\mathbf{y}}}_l).
\label{PCA}
\end{equation}
AP $l$ then passes the real and imaginary part of each element of $\mathcal{P}_l(\mathcal{G}_l({{\mathbf{y}}}_l))$ separately through one quantizer. Therefore, the total number of quantizers in AP $l$ is $2r$ ($r$ pairs). 
\subsection{Uniform Quantization at AP $l$} \label{3B}

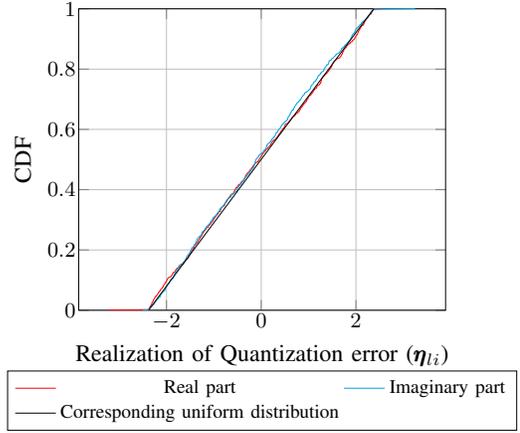
\begin{figure}[h!]
    \centering
    \pgfplotsset{width=7cm,compat=1.18}
\pgfplotsset{every x tick label/.append style={font=\small, yshift=0.5ex},every y tick label/.append style={font=\small, xshift=0.5ex},
every axis legend/.append style={
at={(1.02,1)},
anchor=north west,font=\small
}}

\begin{tikzpicture}[scale=0.9]

\begin{axis}[
grid=major,
ylabel= CDF,
xlabel= Realization of Quantization error ($\pmb{\eta}_{li}$), 
ymin=0,
ymax=1,
legend columns=2,
legend style={nodes={scale=0.9, transform shape},at={(0.5,-0.2)},anchor=north},
]
        
        

        \addplot [red]
        table[x=A,y=C,col sep=comma] {DATA/Data_final_manuscript/cdfinfo_BT_3M.csv};
        \addlegendentry{Real part}

        \addplot [Cerulean]
        table[x=B,y=C,col sep=comma] {DATA/Data_final_manuscript/cdfinfo_BT_3M.csv};
        \addlegendentry{Imaginary part}
      
        \addplot [black]
        table[x=A,y=B,col sep=comma] {DATA/Data_final_manuscript/unifcdfinfo_BT_3M.csv};
        \addlegendentry{Corresponding uniform distribution}




\end{axis}

\end{tikzpicture}
    \caption{Quantization noise of a random pair of quantizers at a random AP. L=5, N=4, K=10, $b_l=3$ and $p=-10\text{dB}$.}
    \label{fig_CDF}
\end{figure}

We use uniform quantization for each element of $\mathcal{P}_l(\mathcal{G}_l({{\mathbf{y}}}_l))$. The bits allocated to each quantizer at AP $l$ is $b_l$. The function $\mathcal{Q}_l$ is an element-wise quantizer. Before quantization, a dither signal vector independent from $\mathcal{P}_l(\mathcal{G}_l({{\mathbf{y}}}_l))$, with zero mean and i.i.d uniformly distributed elements is added to $\mathcal{P}_l(\mathcal{G}_l({{\mathbf{y}}}_l))$. Adding the dither signal before quantization ensures the quantization noise is uniformly distributed and uncorrelated to the input \cite{dither}, which makes the LMMSE estimation of users' signal tractable. 
The dithered and quantized signal vectors are shown respectively as follows:
\begin{equation}
\mathbf{z}_l=\mathcal{P}_l(\mathcal{G}_l({{\mathbf{y}}}_l))+\mathbf{d}_l,
    \label{dither}
\end{equation}
\begin{equation}
\begin{aligned}
    f_{l}(\mathbf{y}_l)&=\mathcal{Q}_l(\mathbf{z}_l+\mathbf{d}_l)\\&\myeq\mathbf{A}_l^{\text{H}}\mathcal{G}_l({{\mathbf{y}}}_l)+\mathbf{d}_l+\pmb{\eta}_l,
    \end{aligned}
    \label{eq_quant}
\end{equation}
where (\ref{eq_quant}) is a result of (\ref{eq_mmse}), (\ref{PCA}) and (\ref{dither}). 
Note that $\operatorname{Re}(\mathbf{d}_l)$ 
and 
$\operatorname{Im}(\mathbf{d}_l)$ 
are also i.i.d uniformly distributed random vectors. The range of the dither signal elements depends on the dynamic range and number of bits of the corresponding quantizers. The dynamic range of the $i^{th}$ pair of quantizers at AP $l$ is $\gamma_{il}$ (quantizers for the real and imaginary part of $i^{th}$ element of dithered signal in (\ref{dither}) have the same dynamic range). If each quantizer at AP $l$ is allocated $b_l$ bits, accordingly, the range of the real and imaginary part of $i^{th}$ element of dither signal vector $\mathbf{d}_{il}$ is as follows:
\begin{equation}
    \Delta_{il}=\frac{2\gamma_{il}}{2^{b_{l}}}.
    \end{equation}
The dither signal vector covariance matrix is then $\mathbf{R}_{\mathbf{d}_l} = \textrm{diag}(\Delta^2_{1l}/6 \, \hdots \Delta^2_{rl}/6)$. We define vector $\pmb{\eta}_l\in\mathbb{C}^{r\times 1}$ as the quantization noise. Based on (\ref{eq_mmse}) and (\ref{eq_quant}):
\begin{equation}
    \mathbf{b}_l=-\mathbf{A}_l^{\text{H}}\mathbf{H}_l\hat{\mathbf{s}}_{l-1}+\mathbf{d}_l+\pmb{\eta}_l.
    \end{equation}
    The elements of $\pmb{\eta}_l$ are also zero-mean uniformly distributed i.i.d 
 random variables with covariance matrix $\mathbf{R}_{\pmb{\eta}_l}=\mathbf{R}_{\mathbf{d}_l}$ and uncorrelated with $\mathcal{P}_l(\mathcal{G}_l({{\mathbf{y}}}_l))$ \cite{Nir}. To validate the claims on distribution, Fig. \ref{fig_CDF} shows the CDF of the quantization noise of the $i^{th}$ pair of quantizers at a random AP $l$ using Monte Carlo simulation compared to a corresponding uniform distribution with range $\Delta_{il}$. Note that the three CDFs almost completely overlap.

For a quantizer to work within the dynamic range, the dynamic range of the quantizer should be some multiple ($\alpha$) of the standard deviation of the input of the quantizer.
Consider the $i^{th}$ element of $\mathbf{z}_l$. The dynamic range of $i^{th}$ pair of quantizers  is calculated as follows:
\begin{equation}
\begin{aligned}
    \gamma_{il}^2&=\alpha^2 \mathbb{E}\{{z}_{li}{z}_{li}^{\text{H}}\}/2\\&=\alpha^2(\mathbb{E}\{\mathcal{P}_l(\mathcal{G}_l({{\mathbf{y}}}_l))_i\mathcal{P}_l(\mathcal{G}_l({{\mathbf{y}}}_l))_i^{\text{H}}\}/2+\frac{\Delta_{li}^2}{12})\\&=\alpha^2( \mathbb{E}\{\mathcal{P}_l(\mathcal{G}_l({{\mathbf{y}}}_l))_i\mathcal{P}_l(\mathcal{G}_l({{\mathbf{y}}}_l))_i^{\text{H}}\}/2+\frac{\gamma^2_{li}}{3\times4^{b_l}})\\&\rightarrow \gamma_{il}=\sqrt{\alpha^2(1-\frac{\alpha^2}{3\times 4^{b_l}})^{-1}\mathbb{E}\{\mathcal{P}_l(\mathcal{G}_l({{\mathbf{y}}}_l))_{i}\mathcal{P}_l(\mathcal{G}_l({{\mathbf{y}}}_l))_{i}^{\text{H}}\}/2},
    \end{aligned}
\end{equation}
where ${z}_{li}$ and $\mathcal{P}_l(\mathcal{G}_l({{\mathbf{y}}}_l))_{i}$ are the $i^{th}$ element of the $\mathbf{z}_l$ and $\mathcal{P}_l(\mathcal{G}_l({{\mathbf{y}}}_l))$, respectively.
\begin{figure*}[ht!]
    \centering
    \includegraphics[width=\textwidth]{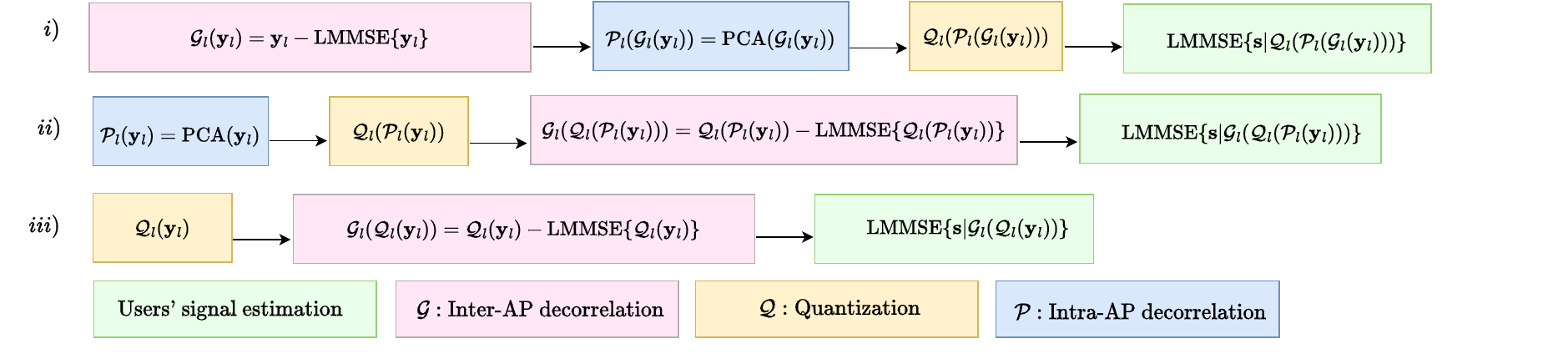}
    \caption{Three possible processing sequences for sequential signal estimation approaches at AP $l$. i) {\bf Option $1$:} Before quantization, AP $l$ first de-correlates its received signal vector from the previous APs vector (\textbf{\textit{inter-AP}} de-correlation) and then uses PCA to de-correlate the dimensions of the resulting vector (\textbf{\textit{intra-AP}} de-correlation). ii) {\bf Option $2$: } AP $l$ only considers de-correlating the dimensions of its own received signal vector (\textbf{\textit{intra-AP}} de-correlation) before quantization, iii) {\bf{Option $3$: }} AP $l$ does not do any de-correlation before quantization and directly quantizes the received signal vector elements (No de-correlation before quantization). Due to space limitations, dithering is not shown in the block diagram.}
    \label{3cases}
\end{figure*}
\subsection{Users' signal estimation at AP $l$}\label{3C}
The resulting quantized vector is sent to the base-band processing unit to update user signal estimates. As mentioned earlier, AP $l$ tries to estimate the unknown part of $\mathbf{s}$ which is $\tilde{\mathbf{s}}_{l-1}$ in (\ref{eq_signal_est_err}).
The MMSE estimate of $\tilde{\mathbf{s}}_{l-1}$ from $\mathcal{Q}_l\{{{\mathbf{z}}}_l\}$ is as follows:
\begin{equation}
\begin{aligned}
    \hat{\tilde{\mathbf{s}}}_{l-1}&=\mathbb{E}\{\tilde{\mathbf{s}}_{l-1}f_{l}({\mathbf{y}}_l)^{\text{H}}\}\}(\mathbb{E}\{f_{l}({\mathbf{y}}_l)f_{l}({\mathbf{y}}_l)^{\text{H}}|\mathbf{H}_l\})^{-1}f_{l}({\mathbf{y}}_l)\\&=\mathbf{C}_{l-1}\mathbf{H}_l^{\text{H}}\mathbf{A}_l\mathbf{R}_{f_{l}({\mathbf{y}}_l)}^{-1}f_{l}({\mathbf{y}}_l),
    \end{aligned}
    \label{local_MMSE}
\end{equation}
where $\mathbf{V}_l=\mathbf{C}_{l-1}\mathbf{H}_l^{\text{H}}\mathbf{A}_l\mathbf{R}_{f_{l}({\mathbf{y}}_l)}^{-1}$. 
 \begin{equation}
\mathbf{R}_{f_{l}({\mathbf{y}}_l)}=\mathbf{A}_l\mathbf{R}_{\mathcal{G}_l({\mathbf{y}}_l)}\mathbf{A}_l^{\text{H}}+\mathbf{R}_{\mathbf{d}_l}+\mathbf{R}_{\pmb{\eta}_l},
 \end{equation}
We show the diagonality of $\mathbf{R}_{\pmb{\eta}_l}$ with the Monte Carlo simulation.
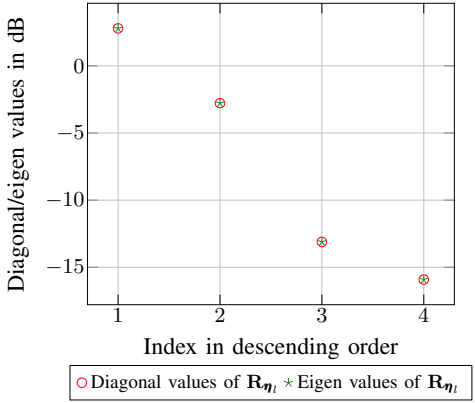
\begin{figure}[h!]
    \centering
    \pgfplotsset{width=7cm,compat=1.18}
\pgfplotsset{every x tick label/.append style={font=\small, yshift=0.5ex},every y tick label/.append style={font=\small, xshift=0.5ex},
every axis legend/.append style={
at={(1.02,1)},
anchor=north west,font=\small
}}

\begin{tikzpicture}[scale=0.9]
\begin{scope}
\begin{axis}[
domain=0:4,
grid=major,
ylabel= Diagonal/eigen values in dB,
xlabel= Index in descending order,
xtick style={color=black},
 xtick={1,2,3,4,5,6,7},
legend columns=2,
legend style={nodes={scale=0.9, transform shape},at={(0.5,-0.2)},anchor=north},
]
        
        

        \addplot [only marks][red, mark=o]
        table[x expr=\coordindex+1,y=A,col sep=comma] {DATA/Data_final_manuscript/diagv.csv};
        \addlegendentry{Diagonal values of $\mathbf{R}_{\pmb{\eta}_l}$}


        \addplot [only marks][OliveGreen, mark=star]
        table[x expr=\coordindex+1,y=A,col sep=comma] {DATA/Data_final_manuscript/eigv.csv};
        \addlegendentry{Eigen values of $\mathbf{R}_{\pmb{\eta}_l}$}



\end{axis}
\end{scope}
\end{tikzpicture}
    \caption{Diagonal vs. eigenvalues of the quantization noise covariance matrix at a random AP. L=5, N=4, K=10, $b_l=3$ and $p=-10\text{dB}$.}
    \label{fig_quant_n_diag}
\end{figure}
Fig. (\ref{fig_quant_n_diag}) shows that the eigenvalues and diagonal elements of the covariance matrix of the quantization noise vector at a randomly selected AP are the same, testifying to the diagonality of the quantization noise covariance matrix. 
The user signal estimates are updated as follows:
\begin{equation}
    \hat{\mathbf{s}}_{l}=\hat{\mathbf{s}}_{l-1}+\hat{\tilde{\mathbf{s}}}_{l-1}.
\end{equation}
Accordingly, the user signal estimation error covariance matrix is updated as follows:
\begin{equation}
\begin{aligned} 
    \mathbf{C}_l&=\mathbb{E}\{(\mathbf{s}-\hat{\mathbf{s}}_{l})(\mathbf{s}-\hat{\mathbf{s}}_{l})^{\text{H}}\}\\&\myeq\mathbb{E}\{(\tilde{\mathbf{s}}_{l-1}-\hat{\tilde{\mathbf{s}}}_{l-1})(\tilde{\mathbf{s}}_{l-1}-\hat{\tilde{\mathbf{s}}}_{l-1})^{\text{H}}\}\\&\myeqb (\mathbf{I}_K-\mathbf{V}_l\mathbf{A}^{\text{H}}_l\mathbf{H}_l)\mathbf{C}_{l-1}.
\end{aligned}
\end{equation}
The updated user signal estimates and error covariance matrix are sent to AP $l+1$.
\subsection{Fronthaul bit rate}\label{3D}
As shown in (\ref{local_MMSE}), the refinement of users' signal in each AP is a multiplication of an $K\times r$ matrix with a vector of dimension $r$ (inner product of the combining vectors with the quantized pre-processed received signal vector). Hence, we have the inner product of two vectors with dimension $r$ for each user. Assume that the real or imaginary part of each element of the combining vector and quantized pre-processed vector has $b_c$ and $b_l$ bits, respectively. For calculating the number of bits of the combining operation, two things should be remembered: 1) the result of the multiplication of two binary numbers (ignoring sign bits) with bit length $b_l$ and $b_c$ can be fully represented using $\rho=b_c+b_l$ bits and 2) the summation of two binary numbers with same bit length $\rho$ can be fully represented using $\rho+1$ bits. For each element of $\hat{\mathbf{s}}_l$, we first multiply $r$ pairs of complex scalars and then add the results together. Then, each element of $\hat{\mathbf{s}}_l$ has up to $b_s=2(b_c+b_l+2r-1)$ bits \cite{rabaey2003digital}. Assume that the user signal estimation error covariance matrix i.e., $\mathbf{C}_l$, needs a total number of $b_e$ bits to be transmitted in each coherence block. As there are $\tau_d$ uplink samples per coherence block, and we assume there are $N_{CB}=B/B_c$ coherence blocks over a time distance of $T_c$ \cite[Chapter~2]{marzetta_larsson_yang_ngo_2016}, the bit rate on a fronthaul link connecting AP $l$ to AP $l-1$ is as follows:
\begin{equation}
    Br_f=\frac{N_{CB}(b_e+2\tau_dK(b_c+b_l+2r-1))}{T_c}.
\end{equation}
So the number of bits to be transmitted on the fronthaul link increases linearly with $b_l$. This means that each AP should try to lower $b_l$ to satisfy fronthaul capacity constraints without compromising performance significantly.

It is also worth mentioning that the quantization for fronthaul can also happen after users' signal estimation. However, this choice will not reduce the hardware complexity of base-band processing. 
Furthermore, in this case, the number of quantizers in the whole network will be $L\times 2K\geq L\times 2r$.
\subsection{Hardware complexity}
As mentioned, aside from affecting fronthaul bit rate, the number of bits of the quantized pre-processed signal vector also affects both the hardware and the time for the multiplication at the base-band processing unit where the multiplication with the combining vector actually happens. For example, for multiplying two numbers with $b_c$ and $b_l$ bits, the number of logical gates linearly increases with $b_l$ \cite{rabaey2003digital}. Hence, having a large $b_l$ increases the size and complexity of the hardware responsible for the combining and user signal estimates refinement in the base-band processing unit of each AP. 

\subsection{Alternative methods}\label{IIIF}
 {\bf Option $1$} processing sequence in an AP, which is elaborated in the Subsections \ref{3A}, \ref{3B} and \ref{3C}, is compared to two alternative approaches: 1) {\bf Option $2$} where the signal received at each AP is only de-correlated \textbf{\textit{intra-AP}}, e.g., by computing the SVD of the received signal vector at the AP and then applying PCA on the received signal vector and 2) {\bf Option $3$} where the received signal vector at the APs passes through quantizers without any \textbf{\textit{inter-AP}} or \textbf{\textit{intra-AP}} de-correlation in advance. The block diagram of the processing sequences corresponding to the three options is shown in Fig. \ref{3cases}.

\section{Numerical experiment} \label{sec4}
In this section, we present the simulation result. A simulation area with a perimeter of $D=500 m$ \cite{Shaik2021} is considered with a total $NL=20$ antennas serving $K=10$ users in the uplink. The simulation parameters are given is table \ref{tab:simPars}. 
\begin{table}[h]
    \centering
        \caption{Simulation Parameters}
    \begin{tabular}{l|r|l|r} 
        Parameter & Value&Parameter & Value \\ \hline \hline
        Bandwidth (B) & 100MHz&Carrier frequency & 2GHz \\ 
        Noise figure & 9dB&Noise variance & -85dBm \\ 
    \end{tabular}
        \label{tab:simPars}
\end{table}
The considered propagation model is the 3GPP Urban Microcell model in \cite{3gpp_PL}, with a large-scale fading coefficient defined as:
\begin{equation}
    \beta_{kl}=-30.5-36.7\log_{10}(\frac{d_{kl}}{1m}),
\end{equation}
where $d_{kl}$ is the distance between user $k$ and AP $l$.
In Fig. \ref{fig_cvdg}, we compare the normalized MSE (NMSE) of user signal estimates when using \textbf{{Option $1$}} processing sequence and the alternative processing sequences as shown in Fig. \ref{3cases}. We observe that with \textbf{{Option $1$}} processing sequence at each AP, the NMSE approaches the level of NMSE of \textbf{\textit{No quantization}} for a relatively smaller number of bits than the other two options. In Fig. \ref{fig_bitr}, while users send the BPSK modulated signal, the bit error rate (BER) of the three aforementioned processing sequence options is bench-marked with the case of \textbf{\textit{No quantization}}. We observe that \textbf{{Option $1$}} shows superior performance compared to the two other alternative options.

\begin{figure}
    \centering
    \pgfplotsset{width=7cm,compat=1.18}
\pgfplotsset{every x tick label/.append style={font=\small, yshift=0.5ex},every y tick label/.append style={font=\small, xshift=0.5ex},
every axis legend/.append style={
at={(1.02,1)},
anchor=north west,font=\small
}}

\begin{tikzpicture}[scale=0.9]
\begin{scope}
\begin{axis}[
domain=0:4,
grid=major,
ylabel= NMSE,
xlabel= Number of bits per quantizer ($b_l$),
xtick style={color=black},
legend columns=4,
legend style={nodes={scale=0.9, transform shape},at={(0.5,-0.2)},anchor=north},
]
        
        

        \addplot [OliveGreen,mark=star]
        table[x=E,y=A,col sep=comma] {DATA/Data_final_manuscript/MSE_K10_L5_N4_p0_1.csv};
        \addlegendentry{No quantization}

         \addplot [Melon, mark=o]
        table[x=E,y=B,col sep=comma] {DATA/Data_final_manuscript/MSE_K10_L5_N4_p0_1.csv};
         \addlegendentry[black]{Option $1$}

        \addplot [Cerulean, mark=square]
        table[x=E,y=C,col sep=comma] {DATA/Data_final_manuscript/MSE_K10_L5_N4_p0_1.csv};
        \addlegendentry{Option $2$}

        \addplot [Maroon, mark=diamond]
        table[x=E,y=D,col sep=comma] {DATA/Data_final_manuscript/MSE_K10_L5_N4_p0_1.csv};
        \addlegendentry{Option $3$}





\end{axis}
\end{scope}
\end{tikzpicture}
    \caption{Average per user NMSE of users signal estimation vs. number of bits of each quantizer. $L=5$, $N=4$, $K=10$, and $p=-10\text{dB}$.}
    \label{fig_cvdg}
\end{figure}
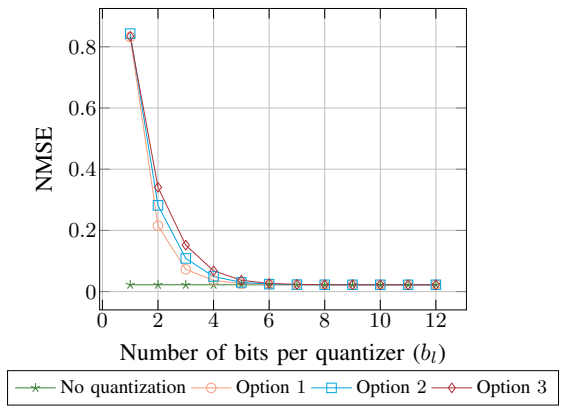
\begin{figure}
    \centering
    \pgfplotsset{width=7cm,compat=1.18}
\pgfplotsset{every x tick label/.append style={font=\small, yshift=0.5ex},every y tick label/.append style={font=\small, xshift=0.5ex},
every axis legend/.append style={
at={(1.02,1)},
anchor=north west,font=\small
}}

\begin{tikzpicture}[scale=0.9]
\begin{axis}
[
grid=major,
ylabel= Bit error rate,
xlabel= user transmit power (p) in dB,
ymode=log,
 xmin=-20,
  xmax=0,
 ymin=0.00001,
 ymax=0.1,
legend columns=4,
legend style={nodes={scale=0.9, transform shape},at={(0.5,-0.2)},anchor=north},
]




        \addplot [OliveGreen,mark=star]
        table[x=E,y=A,col sep=comma] {DATA/Data_final_manuscript/BER_K10_N4_L5_BT3M.csv};
        \addlegendentry{No quantization}
      
        \addplot [Melon, mark=o]
        table[x=E,y=B,col sep=comma] {DATA/Data_final_manuscript/BER_K10_N4_L5_BT3M.csv};
        \addlegendentry{Option $1$}

        \addplot [Cerulean, mark=square]
        table[x=E,y=C,col sep=comma] {DATA/Data_final_manuscript/BER_K10_N4_L5_BT3M.csv};
        \addlegendentry{Option $2$}

        \addplot [Maroon, mark=diamond]
        table[x=E,y=D,col sep=comma] {DATA/Data_final_manuscript/BER_K10_N4_L5_BT3M.csv};
        \addlegendentry{Option $3$}

\end{axis}
\end{tikzpicture}
    \caption{Bit error rate of BPSK modulation vs. user transmit power. $L=5$, $N=4$, $K=10$ and $b_l=3$.}
    \label{fig_bitr}
\end{figure}
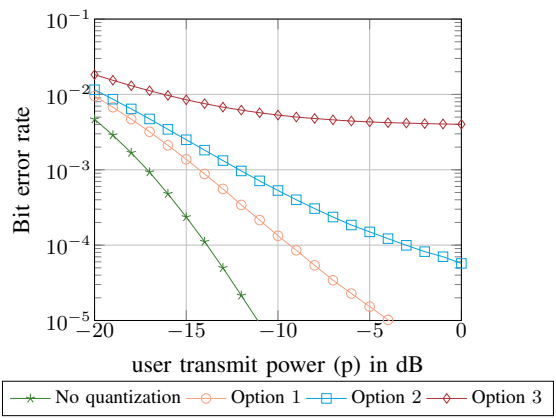


\section{Conclusion} \label{sec5}
In this paper, we consider the efficient quantization of the received signal vector at the APs in a daisy chain cell-free massive MIMO network to 1) reduce the complexity of the local base-band processing unit in each AP, 2) meet the limited capacity requirement of the fronthaul links. 
Furthermore, we demonstrate that the element-wise quantization of the raw received signal vector (\textbf{{Option $3$}}) without \textbf{\textit{intra-AP}} or \textbf{\textit{inter-AP}} de-correlation in advance adversely affects NMSE of user signal estimates and bit error rate performance. On the other hand, de-correlating the local received vector dimensions using PCA before quantization (\textbf{{Option $2$}}) helps to use the bits more efficiently than \textbf{{Option $3$}}, in the considered setup. Ultimately, de-correlating the received signal vectors \textbf{\textit{inter-AP}} (\textbf{{Option $1$}}) before quantization further improves the performance. 

\bibliographystyle{ieeetr}
\bibliography{ref}
\vspace{12pt}


\end{document}